\documentclass[twocolumn,showpacs,preprintnumbers,amsmath,amssymb,prb,superscriptaddress]{revtex4}

\usepackage{mciteplus} \usepackage{graphicx} \usepackage{color}
\usepackage[normalem]{ulem}

\definecolor{darkgreen}{rgb}{0.2, 0.8, 0.2}

\newcommand{\dagind}[2]{#1_{#2}^{\vphantom{\dagger}}}
\newcommand{\intlim}[3]{\int \limits_{#1}^{#2} \text{d} #3}

 \newcommand{\e}{\text{e}}
\newcommand{\ic}{\text{i}}
\begin{document}

\title{Local density of states on a vibrational quantum dot out of equilibrium}
\author{K. F. Albrecht} \affiliation{Physikalisches Institut,Albert--Ludwigs--Universit\"at Freiburg, Hermann-Herder-Str.~3, D--79104 Freiburg, Germany}
\author{A. Martin-Rodero} \affiliation{Departamento de F\'{\i}sica Te\'orica de la Materia Condensada, Universidad Aut\'onoma de Madrid, E28049 Madrid, Spain} 
\author{J. Schachenmayer} \affiliation{JILA, NIST, University of Colorado, 440 UCB, Boulder, CO 80309, USA} 
\author{L. M\"uhlbacher}\affiliation{Physikalisches Institut,Albert--Ludwigs--Universit\"at Freiburg, Hermann-Herder-Str.~3, D--79104 Freiburg, Germany}
\date{\today}
\begin{abstract}
  We calculate the nonequilibrium local density of states on a vibrational quantum dot coupled to two electrodes at $T=0$ using a numerically exact diagrammatic Monte Carlo method. Our focus is on the interplay between the electron-phonon interaction strength and the bias voltage. We find that the spectral density exhibits a significant voltage dependence if the voltage window includes one or more phonon sidebands. A comparison with well-established approximate approaches indicates that this effect could be attributed to the nonequilibrium distribution of the phonons. Moreover, we discuss the long transient dynamics caused by the electron-phonon coupling.
\end{abstract}
\maketitle

Recent experiments in the field of molecular electronics have pointed out the importance of electron-phonon interactions for the charge transport on the nanoscale.\cite{c60,Smit02,Zhitenev02,PhysRevLett.92.206102,Kushmerick04,Liu04,Natelson04,Pasupathy05,Sapmaz06,carbon_nanotube_nature2009} In these experiments, a nanostructure -- such as a single molecule or a carbon nanotube -- is in contact with two electronic leads. Due to the tiny size of the structure, single-electron tunneling processes can cause a transient change of its electronic geometry. This change in combination with intermolecular interactions can couple the electronic to the vibrational degrees of freedom. An important consequence is the appearance of nonlinearities in the current-voltage characteristics and the conductance. \cite{Secker11,Cuevas10,Smit02,Natelson04,Ballmann10,Park02,Zhitenev02} These effects can be associated with the possibility of inelastic processes due to the bias and to the formation of phonon sidebands in the excitation spectrum. \cite{PhysRevB.50.5528,Flensberg2003,PhysRevB.69.245302,carmina2010}

From a theoretical perspective, such a quantum dot setup can be described by the Anderson-Holstein model \cite{Holstein1959,Hewson2002}. To its full extent, this model accounts for a tunneling coupling between the quantum dot and electronic reservoirs, a linear coupling of the electrons occupying the quantum dot to phonons, as well as an on-site Coulomb interaction. In this paper, we are mainly interested in the effects of the molecule's vibration on the charge transport through the quantum dot so that it is expedient to consider a single vibrational mode and spinless electrons. 
Therefore, the model can be simplified to account for a single electronic level which is linearly coupled to a local phonon, whereas the electron-electron coupling is disregarded.
In the framework of this spinless Anderson-Holstein model, a lot of progress has been made, offering a deep insight into the physics caused by the coupling of the electron and the phonon (see, e.g.,\ Refs.~[\onlinecite{PhysRevB.50.5528,Flensberg2003,PhysRevB.69.245302,PhysRevB.74.075326,alvaro2008,carmina2010,PhysRevB.84.125131}]). Besides approximative approaches numerically exact methods have (recently) become possible for a nonequilibrium situation of a vibrational quantum dot (see, e.g.,\ Refs.~[\onlinecite{Lothar_diagMC,Wang09,PhysRevB.85.121408,PhysRevB.87.195112,PI_2013}]).

A central quantity to describe the nonequilibrium transport through the quantum dot is its local density of states (LDOS). Single-particle observables such as the current or the dot occupation can be directly derived from it\cite{Meir92}. In equilibrium, the spectral density is well understood (see Refs.~[\onlinecite{0022-3719-13-24-011,Hewson2002,carmina2009}] and references therein). Moreover, the close connection between nanomechanical vibrations of the quantum dot and the sidebands has been confirmed. The interplay between nanomechanical vibrations and a finite bias voltage, however, still remains a challenging task outside certain limiting cases. \cite{Galperin06,PhysRevB.87.195112,1367-2630-16-2-023007}

In this paper, we address this problem in a numerically exact way by using the diagrammatic Monte Carlo method \cite{Lothar_diagMC,Schiro09,Werner2011} (diagMC). For this purpose, we use a two-terminal setup with an auxiliary electrode. This allows for an exact study of the LDOS on a quantum dot coupled to two electrodes in the limit of a vanishing coupling to the auxiliary lead.\cite{sun_guo_third_terminal2001,lebanon_schiller_third_terminal2001,Lothar_spectral_density} Throughout this paper we consider the deep quantum limit at $T=0$. 

For a thorough discussion of the numerical results, we use an interpolative self-energy approximation (ISA), in which it is possible to include electron-phonon interactions \cite{alvaro2008,carmina2010}, and the well-established single particle approximation \cite{PhysRevB.50.5528,PhysRevB.66.085311,PhysRevB.66.075303,Flensberg2003,PhysRevB.76.033417} (SPA). Although these approximations rely on completely different approximation schemes, one uses the underlying common assumption that the phonons are described via an equilibrium distribution. Consequently, effects due to a nonequilibrium distribution of phonons can be clearly identified by comparing these approximations to the numerically exact results.

The structure of the paper is as follows: In Section~\ref{model} we introduce the Anderson-Holstein model. In Section~\ref{measuring_the_spectral_function} we show how the diagMC can be used to calculate the LDOS for a quantum dot with an electron-phonon interaction by adapting the approach of Ref.~[\onlinecite{Lothar_spectral_density}]. Moreover, the ISA, SPA and certain limiting cases are briefly summarized. The results for weak electron-phonon couplings are presented in Section~\ref{electronic_regime}. The moderate polaronic regime is addressed in Section~\ref{polaronic_regime}.

\section{The model} 
\label{model}

In our discussion, we consider a molecular quantum dot connected by a tunneling coupling to a left (L) and right (R) electrode. For such a system, a single-electron charging of the quantum dot can cause a nanomechanical vibration of the molecule\cite{c60}. A reasonable model Hamiltonian for this situation is provided by the spinless Anderson-Holstein model \cite{Glazman1988,Galperin07} (throughout this paper we use units with $\hbar=e=k_{\text{B}}=1$):

\begin{align}
  \nonumber
  H 
  &=
  \dagind{\sum}{k \in \alpha} 
  \left(
    \dagind{\epsilon}{\alpha k} 
    -
    \mu_{\alpha}
  \right)
  a^{\dagger}_{\alpha k} \dagind{a}{\alpha k}
  +
  \sum_{k \in \alpha} 
  \gamma_{\alpha} 
  \left(
    a^{\dagger}_{\alpha k} 
    d 
    + 
    d^{\dagger}
    a^{\vphantom{\dagger}}_{\alpha k} 
  \right) 
  \\
  & \quad
  +
  \epsilon_{\text{D}} d^\dagger d
  +
  \lambda d^{\dagger} d \left( b + b^{\dagger} \right)
  +
  \omega_0 b^{\dagger} b
  \,.
  \label{Eq:Anderson-Holstein_Hamiltonian}
\end{align}

The quantum dot is modeled by a single electronic energy level at $\epsilon_{\text{D}}$. $d^{\dagger}$ ($d$) is the electron creation (annihilation) operator on the dot. $\alpha = \text{L}$ denotes the left and $\alpha = \text{R}$ the right electrode. The electronic creation (annihilation) operator on electrode $\alpha$ at energy level $\epsilon_{\alpha k}$ is denoted by $a^{\dagger}_{\alpha k}$ ($a_{\alpha k}$). The bias voltage, $V=\mu_{\text{L}} - \mu_{\text{R}}$, is defined as the difference between the two chemical potentials $\mu_{\alpha}$ of the respective electrode. This quantity is assumed to be constant for all times.  $\gamma_{\alpha}$ are the tunneling amplitudes. The tunneling rates in the absence of manybody effects are given by $\Gamma_{\alpha} = 2 \pi \rho_{\alpha} | \gamma_{\alpha}| ^2 $, where $\rho_{\alpha}$ is the density of states of lead $\alpha$, which is assumed to be a flat band. The linear coupling of the vibrational mode to the electronic degrees of freedom is described by a single phonon mode with frequency $\omega_0$. $b^{\dagger}$ and $b$ are the phonon creation and annihilation operators, respectively. $\lambda$ denotes the electron-phonon coupling constant.

In the subsequent discussion, we study the deep quantum limit at $T=0$. Moreover, a symmetric setup is assumed where $\Gamma_{\text{L}}=\Gamma_{\text{R}}=\Gamma/2$, $\mu_{\text{L}}=-\mu_{\text{R}}=V/2$ and we consider the particle-hole symmetric case, $\tilde{\epsilon}_{\text{D}}=0$. $\tilde{\epsilon}_{\text{D}} = \epsilon_{\text{D}} - \lambda^2/\omega_0$ is the polaron-shifted energy level of the quantum dot. This parameter regime is very interesting since the steady-state dot occupation is always $\langle n \rangle = 0.5$ due to the symmetry of the setup. Consequently, any bias dependence of the LDOS can only be caused by the electron-phonon interaction and not the electronic
occupation of the quantum dot itself.

\begin{figure}
  \begin{center}
    \includegraphics[width=0.475\textwidth]{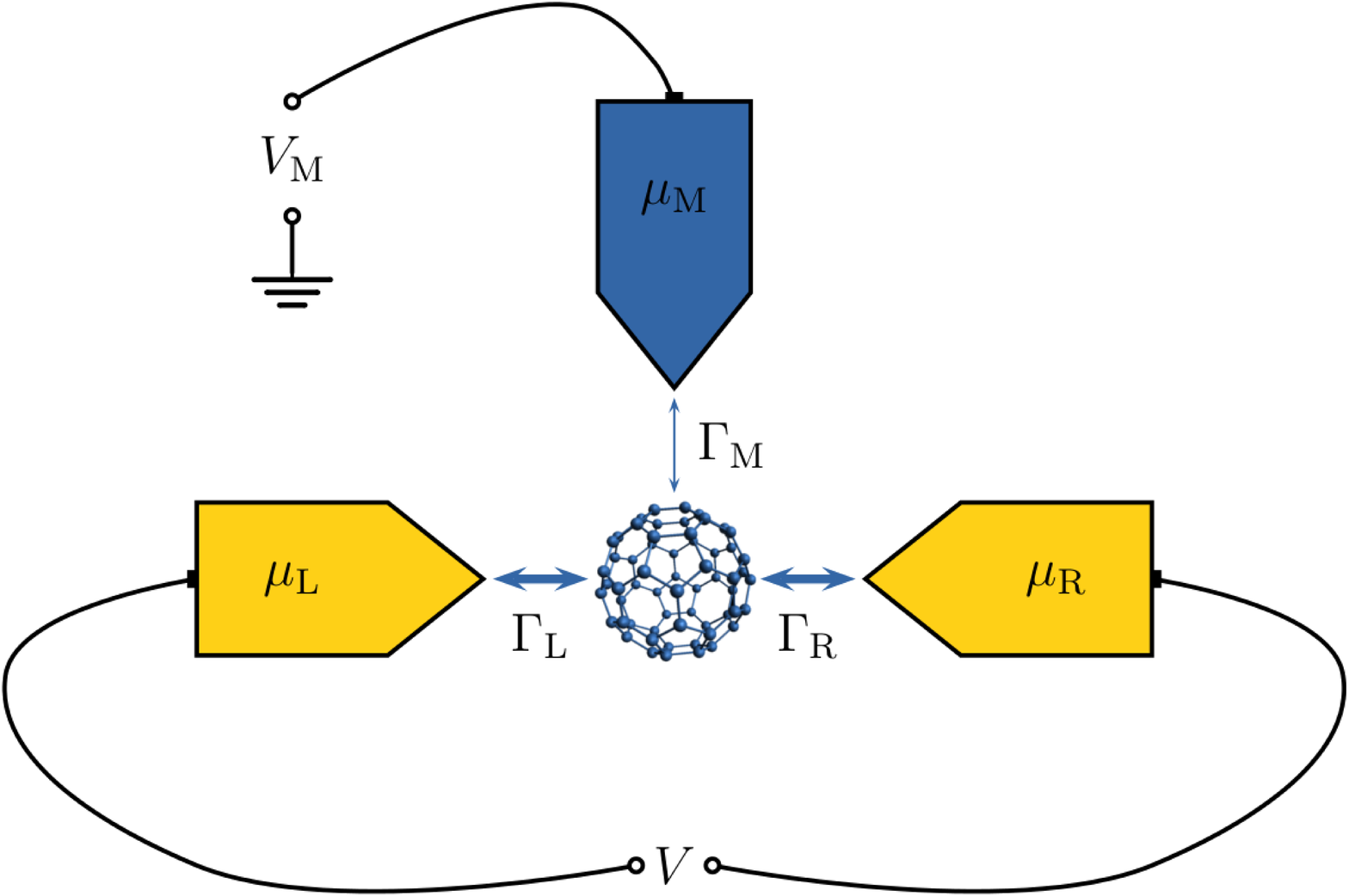}
    \caption{Sketch of the three-terminal setup with a weak coupling of the auxiliary lead to the quantum dot $\Gamma_{\text{M}} \ll \Gamma_{\text{L}}, \Gamma_{\text{R}}$. The bias voltage $V_{\text{M}}$ can be tuned in order to access the desired energy in the spectral density of the quantum dot. $\mu_{\text{M}}$ is the respective chemical potential of the auxiliary electrode. \label{Fig:sketch_third_electrode}}
  \end{center}
\end{figure}
\section{Approaches for the spectral density of a vibrational quantum dot}\label{measuring_the_spectral_function}

\subsection{Diagrammatic Monte Carlo simulation method}

Despite recent progress in developing analytical approaches for the Anderson-Holstein model, e.g., by means of diagrammatic resummation schemes \cite{ruben,PhysRevB.83.085401,apta}, a complete solution outside certain limiting cases is currently unknown. In order to calculate the spectral density without having to rely on methods which involve intrinsic approximations, numerical methods are needed (see, e.g., Refs.~[\onlinecite{Lothar_diagMC,Wang09,PhysRevB.85.121408,PhysRevB.87.195112,PI_2013}]). A suitable approach to access regimes of arbitrary voltage, electron-phonon interaction, and dot-lead coupling strength is the numerically exact diagMC method\cite{Lothar_diagMC,Werner2009,Schiro09,Werner2011}, which is able to simulate finite temperatures as well as $T=0$.

 In the subsequent discussion we use a similar approach to that of Ref.~[\onlinecite{Lothar_spectral_density}], where the diagMC has been used to calculate the LDOS for the Anderson impurity model. Here, we briefly summarize this approach and show how it can be adapted for the case of a local electron-phonon interaction on the quantum dot.

Following the lines of Refs.~[\onlinecite{lebanon_schiller_third_terminal2001,sun_guo_third_terminal2001}], the spectral density of a two-lead quantum dot can be calculated exactly using an auxiliary lead at chemical potential $\mu_{\text{M}}$ with a vanishing dot-lead coupling (see Fig.~\ref{Fig:sketch_third_electrode} for a sketch of the setup). Their basic idea is to generalize the Meir-Wingreen formula for the current \cite{Meir92} to a three-terminal setup to obtain

\begin{align}
  \nonumber 
  I_{\text{M}}
  &=
  \frac{\Gamma_{\text{M}}}{\Gamma+\Gamma_{\text{M}}}
  \int 
  \text{d} \omega
  \rho_{\text{D}}(\omega)
  \\
  & \qquad
  \times
  \left[
    \Gamma
    f_{\text{M}}(\omega)
    -
    \Gamma_{\text{L}}
    f_{\text{L}}(\omega)
    -
    \Gamma_{\text{R}}
    f_{\text{R}}(\omega)
  \right]
  \label{eq:current_third_lead}
  \,,
\end{align}

where the subscript M denotes the auxiliary lead and $\rho_{\text{D}}(\omega)$ is the LDOS of the quantum dot. $f_{\alpha}(\omega)=1/(e^{\beta(\omega-\mu_{\alpha})}+1)$ denotes the Fermi function of lead $\alpha$, where $\beta$ is the inverse temperature.

In the limit of a vanishing tunneling coupling of the auxiliary electrode, i.e.\ $\Gamma_{\text{M}} \to 0$, one obtains the thermally broadened LDOS of the two-electrodes setup by deriving Eq.~(\ref{eq:current_third_lead}) with respect to the chemical potential of the auxiliary lead:\cite{lebanon_schiller_third_terminal2001}

\begin{align}
  \lim \limits_{\Gamma_{\text{M}} \to 0}
  \Gamma_{\text{M}}^{-1} 
  \frac{\partial I_{\text{M}}}{\partial \mu_{\text{M}}}
  &=
  \int 
  \text{d} \omega
  \rho_{\text{D}}(\omega)
  \frac{\partial f_{\text{M}}(\omega)}{\partial \mu_{\text{M}}}
  \label{eq:conductance_third_terminal_spectral_function}
  \,.
\end{align}

In the deep quantum limit, i.e.\ at $T=0$, the derivative of the Fermi function in Eq.~(\ref{eq:conductance_third_terminal_spectral_function}) becomes a delta distribution so that the exact LDOS of the two-electrodes setup is obtained and the thermal broadening vanishes\cite{Lothar_spectral_density}:

\begin{align} 
  \rho_{\text{D}}(\mu_{\text{M}}) 
  &= 
  \lim \limits_{\Gamma_{\text{M}} \to 0}
  \Gamma_{\text{M}}^{-1} 
  \frac{\partial I_{\text{M}}}{\partial \mu_{\text{M}}}
\label{eq:steady_state_lebanon_schiller} 
\,.
\end{align}

A convenient way to evaluate Eq.~(\ref{eq:steady_state_lebanon_schiller}) for the Anderson-Holstein model is to use the diagrammatic expansion in the tunnel coupling\cite{Lothar_diagMC,Werner2009,Werner2011}. This expansion allows for a complete decoupling of the influence of the leads to the dot, denoted by $\mathcal{L}_{\text{M}}(\vec{s}_n)$, from the phononic one including the dot's energy level, denoted by $\mathcal{G}(\vec{s}_n)$. Therefore, with the use of Eq.~(\ref{eq:steady_state_lebanon_schiller}) one obtains the transient which establishes the LDOS starting from an initially decoupled preparation:

\begin{align}
  \rho_{\text{D}}(\mu_{\text{M}})
  &=
  2
  \lim \limits_{t \to \infty}
  \lim \limits_{\Gamma_{\text{M}} \to 0}
  \Gamma_{\text{M}}^{-1}
  \sum \limits_{n=1}^{\infty} 
  \left(
    -1
  \right)^{n}
  \intlim{0}{t}{\vec{s}_n
  \nonumber}
  \\
  & \qquad\qquad
  \times
  \frac{\partial}{\partial \mu_{\text{M}}}
  \text{Re}
  \left \lbrace
    \mathcal{L}_{\text{M}}(\vec{s}_n)
    \mathcal{G}(\vec{s}_n)
  \right \rbrace
  \label{eq:diagrammatic_expansion}
  \,.
\end{align}

We used the abbreviation

\begin{align} 
  \int \limits_0^{t} 
  \text{d} \vec{s}_n 
  \equiv 
  \int \limits_0^{t} 
  \text{d}s_{2n} 
  \int \limits_0^{s_{2n}} 
  \text{d}s_{2n-1}
  \cdots 
  \int \limits_0^{s_2} 
  \text{d}s_1 
  \,,
\end{align}
where $\vec{s}_n=\{s_1,s_2,\dots,s_{2n}\}$ is the time-ordered sequence of $2n$ tunneling times $s_j$. 

While in Ref.~[\onlinecite{Lothar_spectral_density}] $\mathcal{G}$ accounts for a Coulomb on-site interaction, in Eq.~(\ref{eq:diagrammatic_expansion}) it provides the influence of the electron-phonon interaction given by \cite{Lothar_diagMC}

\begin{align} 
  \mathcal{G}(\vec{s}_n)
  &= 
  \mathcal{F}[\vec{s}_n] \ 
  \e^{
    \ic 
    \tilde{\epsilon}_{\text{D}} 
    \left( 
      s_1 
      - 
      s_2 
      + 
      s_3 
      -
      \cdots 
    \right) 
  }
  \label{eq:phonon_influence} \,,
\end{align}
where an initially empty quantum dot is considered. ${\mathcal F}[\vec{s}_n]$ denotes the Feynman-Vernon influence functional\cite{Feynman63}:
\begin{align} 
  {\mathcal F}[\vec{s}_n]
  &= 
  \exp 
  \left \lbrace 
    -
    \int_{\mathcal{C}} \text{d}s_1 \int_{\mathcal{C}: s_2 < s_1} \!
    \text{d}s_2 q(s_1) L(s_1-s_2) q(s_2) 
  \right \rbrace \,.
\end{align}

The integrations are performed on the Keldysh contour $\mathcal{C}: 0 \to t \to 0$.  $q(s)$ denotes the occupation of the quantum dot at time $s$ which is fully determined by the initial condition of the quantum dot and the position as well as the number of the tunneling events given by $\vec{s}_n$. For the considered single phonon mode, the bath autocorrelation function is given by

\begin{align}
  L(s)
  &= 
  \frac{
    \lambda^2
  }{
    \omega_0
  } 
  \left[ 
    \cos(\omega_0s) 
    - 
    \ic 
    \sin(\omega_0 s) 
  \right] 
  \,.
\end{align}

We would like to emphasize that the only assumption for this approach is that the quantum dot is initially decoupled from the leads. In detail, right before the coupling, the leads and the quantum dot are considered to be in their respective thermal equilibrium. The nonequilibrium aspect of the system enters via the coupling of the leads to the quantum dot at $t=0$. This causes some transient dynamics until a nonequilibrium steady state is reached.

Since $\mathcal{G}(\vec{s}_N)$ is independent of the leads, the derivative with respect to the chemical potential of the auxiliary lead in Eq.~(\ref{eq:diagrammatic_expansion}) only acts on $\mathcal{L}_{\text{M}}(\vec{s}_N)$, which is a determinant of a matrix consisting of lesser and greater self-energies of the decoupled leads. Calculating this derivative, one obtains at $T=0$: \cite{Lothar_spectral_density}

\begin{align} 
  \lim \limits_{\Gamma_{\text{M}} \to 0}
  \frac{1}{\Gamma_{\text{M}}}
  \frac{\partial}{\partial \mu_{\text{M}}}
  \mathcal{L}_{\text{M}}(\vec{s}_n) 
  &= 
  \ic^n 
  \det
  (
  \mathcal{S}^{\text{M}}(\vec{s}_n)
  )
  \label{eq:leads_influence} 
  \,,
\end{align}
where
\begin{align} 
  \mathcal{S}^{\text{M}}_{j,k}(\vec{s}_n) 
  &= 
  \left \lbrace
    \begin{array}{ll} 
      \Sigma^{<}(s_{2k-1},s_{2j}) &\mbox{, for } j \le k 
      \\ 
      \Sigma^{>}(s_{2k-1},s_{2j}) &\mbox{, for } j > k \\
      \frac{\ic}{2\pi} \e^{-\ic \mu_{\text{M}} (s_{2k-1}-s_{2j})} &\mbox{, if } (s_{2k-1} \lor s_{2j}) = t
    \end{array} \right.
  \label{eq:leads_determinant} 
  \,.
\end{align}

We would like to emphasize that the limit $\Gamma_{\text{M}} \to 0$ has been performed analytically so that the auxiliary lead is not influencing the two-terminal quantum dot. Consequently, the stationary limit of Eq.~(\ref{eq:diagrammatic_expansion}) is the exact LDOS of the two-terminal setup. We would like to note that the steady state is defined with respect to the reduced dynamics of the quantum dot. Therefore, the complete system is in nonequilibrium even though a time-independent steady state for observables on the quantum dot is reached.

Therefore, for a given sequence of tunneling events, $\vec{s}_n$, it is straightforward to calculate the influence of the leads on the dot as well as the phononic influence without any approximation. The summation and integration over all possible tunneling events in Eq.~(\ref{eq:diagrammatic_expansion}) can be done conveniently in a numerical exact manner by using Monte Carlo sampling \cite{Lothar_diagMC,Schiro09,Werner2011}. Using this method, the only occurring error is a controllable statistical one. We note that whenever no error bar is visible in the subsequent figures, the error is smaller than the symbol size.

In the following we will consider two different coupling procedures of the leads to the dot at $t=0$:  A sudden, and a smooth switch-on (for details of the coupling procedure see, e.g., Ref.~[\onlinecite{Lothar_spectral_density}]). In addition, we truncate the leads' density of states at a value $\pm \epsilon_{\text{c}}$. The reason for this sharp cutoff is that an instantaneous coupling of the electrodes to the quantum dot can lead to excitations in the leads, which are arbitrarily high in energy at $t=0$. \cite{Schmidt2008} These short-living excitations are not only unphysical but also make a numerical evaluation using diagMC unfeasible. For our results, the cutoff is chosen to be the largest energy scale in the system so that a further increase of $\epsilon_{\text{c}}$ does not change our results for times $t \gtrsim \epsilon_{\text{c}}^{-1}$. \cite{Schmidt2008}

\subsection{Limiting cases}\label{limiting_cases}

In this section, we briefly discuss two limiting cases, which can be solved analytically. The first one is the absence of electron-phonon interactions, $\lambda/\Gamma \to 0$. Here, it is straightforward to see that the LDOS is independent of the applied bias voltage \cite{PhysRev.124.41}:

\begin{align}
  \rho_{\text{D}}(\omega)
  &=
  \frac{1}{2 \pi}
  \frac{
    \Gamma
  }{
    \left(
      \omega
      -
      \epsilon_{\text{D}}
    \right)^2
    +
    \left(
      \Gamma
      /
      2
    \right)^2
  }
  \label{eq:spectral_function_free}
  \,.
\end{align}
Considering a weakly coupled phonon mode, the electron-phonon coupling can be treated perturbatively \cite{Flensberg2003,PhysRevB.74.075326,PhysRevLett.103.136601,PhysRevB.80.041307,PhysRevB.80.041309,Riwar2009}. Consequently, Eq.~(\ref{eq:spectral_function_free}) indicates that in the perturbative regime no, or only a weak voltage dependence in the spectral density is expected.

Similar arguments hold for the atomic limit, where the electron-phonon coupling becomes very large, $\Gamma/\lambda \to 0$. In this case, the LDOS is given by sharp delta-peaks at multiples of the phonon frequency \cite{Mahan1991}
\begin{align}
  \rho_{\text{D}}
  (\omega)
  &=
  e^{-g}
  \sum_{k=0}^{\infty}
  \frac{g^k}{k!}
  \left \lbrace
    \left[
      1- \langle n_{\text{D}} \rangle
    \right]
    \delta(\omega-\tilde{\epsilon}_{\text{D}}-k \omega_0)
  \right.
  \nonumber
  \\
  & \quad
  \left.
    +
    \langle n_{\text{D}} \rangle
    \delta(\omega-\tilde{\epsilon}_{\text{D}}+k \omega_0)
  \right \rbrace
  \label{eq:spectral_function_atomic_limit}
  \,,
\end{align}
where $\langle n_{\text{D}} \rangle$ is the charge on the quantum dot, and $g=(\lambda/\omega_0)^2$ is the dimensionless electron-phonon coupling strength. Since in this limit the leads are decoupled from the quantum dot, the LDOS cannot depend on the bias voltage.

To summarize these considerations, it is clear that in the weak as well as in the strong coupling limit, a possible voltage dependence of the LDOS can only be weak. Therefore, the nonequilibrium LDOS outside these limiting cases is expected to be the most interesting one.

\subsection{Approximative approaches}

Finally, we  discuss two important and often used approximative approaches. 

A very popular approximate scheme is the SPA \cite{PhysRevB.50.5528,PhysRevB.66.085311,PhysRevB.66.075303,Flensberg2003,PhysRevB.76.033417,carmina2009}, where the electrons are decoupled from the phonons. For this approach the LDOS at temperature $T$ can be written as

\begin{align}
\rho_{\text{D}}(\omega)
  &=
  \frac{\Gamma}{2\pi}
  e^{-g(2n_{\text{B}}+1)}
  \sum \limits_{k=-\infty}^{\infty}
  I_k[2g\sqrt{(n_{\text{B}}+1)n_{\text{B}}}]e^{k\beta\omega_0/2}
  \nonumber
  \\
  &\qquad
  \times
  \left[
    \frac{
      1- \langle n_{\text{D}} \rangle
    }{
      \left(
        \omega
        -
        \tilde{\epsilon}_{\text{D}}
        -
        k \omega_0
      \right)^2
      +
      \left(
        \Gamma/2
      \right)^2
    }
  \right.
  \nonumber
  \\
  &\qquad \quad
  \left.
    +
    \frac{
      \langle
      n_{\text{D}}
      \rangle
    }{
      \left(
        \omega
        -
        \tilde{\epsilon}_{\text{D}}
        +
        k \omega_0
      \right)^2
      +
      \left(
        \Gamma/2
      \right)^2
    }
  \right]
  \label{eq:spectral_function_spa}
  \,,
\end{align}

where the charge on the quantum dot, $\langle n_{\text{D}} \rangle$, has to be calculated self-consistently. $I_k$ is the  modified Bessel function, $\beta=1/T$ and $n_{\text{B}}=1/(e^{\beta\omega_0}-1)$. The simple structure of the SPA allows for a straightforward evaluation. Moreover, it provides good results if the correlations between electrons and phonons are weak, or if the quantum dot is either empty or occupied. Furthermore, the atomic limit as well as the case of absent phonons is recovered. Outside these limiting cases, methods are needed which go beyond the simple SPA decoupling scheme. A well-established method, which is known to provide reasonable results for a broad range of parameters is the ISA \cite{alvaro2008}. The basic idea is to perform a functional interpolation of the self-energies from the weak to the strong coupling regime. This scheme was originally derived for the Anderson impurity model \cite{ISA_Anderson_1,ISA_Anderson_2} and has been extended and widely used in different systems: multilevel quantum dots \cite{ISA_ML_QD}, out-of-equilibrium transport through a single level \cite{ISA_SL_1,ISA_SL_2}, and in dynamical mean-field theory\cite{ISA_DMFT_1,ISA_DMFT_2} to analyze the Mott transition in Hubbard-like models. For the nonequilibrium Anderson-Holstein model this approach provides accurate results beyond perturbation theory or SPA \cite{carmina2010}. 

Using the ISA and the SPA, the effects of the electron-phonon interaction on the charge transport can be discussed qualitatively as long as their basic underlying assumption is fulfilled: an equilibrium distribution of the phonons. Since both methods cover a broad range of parameters this in turn implies that if the physics is qualitatively not covered by either of these methods, there is a strong indication that the phonons no longer obey an equilibrium distribution.

\section{Weakly coupled phonon mode}
\label{electronic_regime}

We start the discussion of our results by considering the weak-coupling regime with $\lambda=\Gamma$, and a rather large phonon frequency $\omega_0=4\Gamma$. The corresponding polaronic self-energy can be determined to be $\Lambda_{\text{pol}}=\lambda^2 / \omega_0 = \Gamma/4$. Since $\Lambda_{\text{pol}} < \Gamma$ the formation of a polaron is a relatively rare event and the electrons are thus weakly coupled to the phonons.

\begin{figure}[tb]
  \begin{center}
    \includegraphics[width=0.475\textwidth]{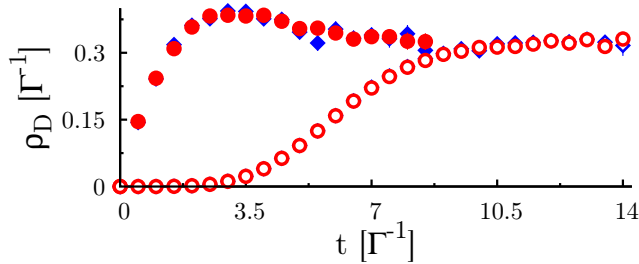}
    \caption{Time-dependent {diagMC results for $\rho_{\text{D}}(\omega, t)$ with $\lambda=\Gamma$, $\omega_0=4\Gamma$ for $V=2\Gamma$ and $\omega=0.5\Gamma$ (red circles) as well as $\omega = -0.5\Gamma$ (blue diamonds)}. Empty symbols denote results for a switch-on time of $\tau_{\text{sw}}= 10 \Gamma^{-1}$. The bandwidth is $2\epsilon_{\text{c}}=8 \Gamma$.\label{Fig:spectral_function_weak_transient_mu0p5}}
  \end{center}
\end{figure}

In Fig.~\ref{Fig:spectral_function_weak_transient_mu0p5} the transients which establish the spectral density for $V=2\Gamma$ at $\omega=\pm 0.5\Gamma$ are shown. An instantaneous coupling to the electrodes leads in this case to an overshooting, with the steady state being approached monotonically. The relevant timescales for the dynamics can be estimated to be $\mathcal{O} \left( \Gamma^{-1} \right)$. A smooth coupling of the quantum dot to the electrodes \cite{Lothar_spectral_density} establishes the steady state adiabatically.

Since the particle-hole symmetric case is considered, the spectral density in the stationary limit must be symmetric with respect to $\omega=0$. In Fig.~\ref{Fig:spectral_function_weak_transient_mu0p5} it can be seen that in the weak-coupling case this property is well fulfilled even in the transient regime.

\begin{figure}[tb]
  \begin{center}
    \includegraphics[width=0.475\textwidth]{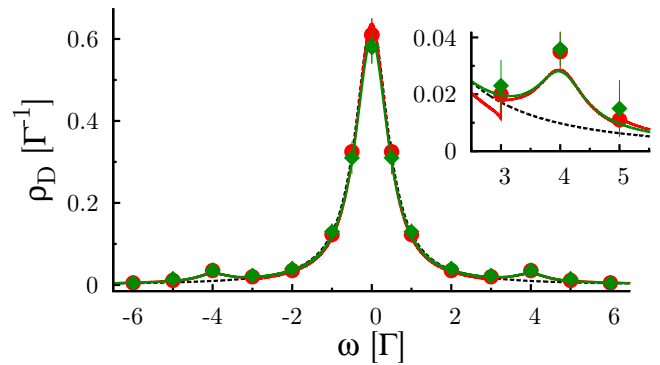}
    \caption{LDOS at zero temperature for a weakly coupled phonon mode with $\lambda=\Gamma$, $\omega_0=4\Gamma$. Red indicates $V=2\Gamma$ and green $V=10\Gamma$. The straight lines are the ISA, circles and diamonds denote the diagMC results. The LDOS in the absence of phonons ($\lambda=0$), given by Eq.~(\ref{eq:spectral_function_free}), is shown as a black dashed line. The inset shows a zoom into the region of the first phonon sideband at $\omega = \omega_0 = 4 \Gamma$. \label{Fig:spectral_function_weak}}
  \end{center}
\end{figure}

The resulting LDOS in the frequency domain is shown in Fig.~\ref{Fig:spectral_function_weak} for two different voltages: $V=2\Gamma$ and $V=10\Gamma$. Comparing results from ISA to the results extracted from the time-dependent diagMC calculations, we find excellent agreement.

The overall shape of the spectral density is very similar for both voltages. A comparison with the results in the absence of phonons ($\lambda=0$) reveals that the height of the central peak remains almost unchanged in the low-voltage regime. The ISA reveals a slight decrease of the central peak when increasing the bias voltage. Since this decrease is small, the Friedel-Langreth sum rule \cite{Friedel1951,PhysRev.150.516}, which pins the height of the central peak to $ \rho_{\text{D}}(0) = 2/(\pi \Gamma)$, \cite{carmina2010} provides a good approximation also for the nonequilibrium situation. 

Small sidebands at multiples of the phonon frequency are observed, which are independent of the voltage within the accuracy of the results extracted from diagMC. The ISA reveals features at $|\omega|=k \omega_0 \pm V/2$, with $k$ being an integer, where the LDOS is changing rapidly (see inset of Fig.~\ref{Fig:spectral_function_weak}). These features can be attributed to inelastic electron tunneling processes as predicted by different theoretical approaches \cite{carmina2010,ruben} to appear both in the spectral density and conductance. For the considered nonequilibrium spectral density with a particle-hole symmetric setup they appear at the condition $|\omega|=k \omega_0 \pm V/2$. For large $k$ the effect can be tiny due to the large amount of phonons involved.

We conclude that the voltage dependence of the spectral density for a weakly coupled phonon mode is very small. This observation is in excellent agreement with our discussion of the limiting cases given in Section~\ref{limiting_cases}. An important consequence for future theoretical approaches is that in this regime it is sufficient to solve the nonequilibrium problem by calculating the LDOS using equilibrium theory. The nonequilibrium aspect of the system enters only via the integration limits for single-particle observables such as the dot occupation or the current.

\section{Spectral density in the moderate polaronic regime}
\label{polaronic_regime}

In the polaronic regime, the formation time of a polaron is shorter than the average occupation time of the electron on the quantum dot. Consequently, the formation of a polaron becomes likely so that pronounced phonon sidebands are expected. The corresponding parameter regime can be determined to be $\Lambda_{\text{pol}} > \Gamma$.

The focus of the subsequent discussion is on the voltage dependence of the spectral density. Our approach to distinguish different voltage regimes is by considering the number of phonon sidebands, which are included in the voltage window: If only the central transport channel is between the two chemical potentials, that is $V \lesssim \omega_0$, the low-voltage regime is realized. For voltage windows including one or more sidebands, we expect that nonequilibrium aspects are most pronounced.

\subsection{Low-voltage regime}
\label{spectral_density_v2}

Important references in the low-voltage regime are equilibrium results such as the Friedel-Langreth sum rule, which is fulfilled at $V=0$ independent of the electron-phonon interaction strength\cite{carmina2009}. Since phonon sidebands form when increasing the electron-phonon coupling \cite{0022-3719-13-24-011,Hewson2002,Flensberg2003,carmina2009}, and the norm of the spectral density needs to be preserved, the central peak must be phonon-narrowed. Such a narrowing can be determined to be $\tilde{\Gamma} \simeq \Gamma \e^{-g}$. \cite{0022-3719-13-24-011} This narrowing of the central peak for strong electron-phonon couplings is the origin of the Franck-Condon blockade effect discussed in detail in Ref.~[\onlinecite{Koch06}]

Regarding the nonequilibrium problem, in Ref.~[\onlinecite{apta}] it was shown via an approximative study that such a narrowing of the resonances can increase the timescales relevant for the charge transport up to
\begin{align}
  \tau_{\text{pol}}
  &=
  \exp
  \left[
    \left(
      \lambda/\omega_0
    \right)^2
  \right]
  \Gamma^{-1}
  =
  \tilde{\Gamma}^{-1}
  \label{eq:narrowing_timescales}
  \,.
\end{align}
Correspondingly, e.g., the transients for establishing the central peak of the spectral density at $\omega=0$ can be estimated to be:
\begin{align}
  \rho_{\text{D}}(\omega=0,t)
  &=
  \frac{
    2
  }{
    \pi \Gamma
  }
  \left(
    1
    -
    e^{- t \tilde{\Gamma}/2}
  \right)
  \label{eq:long_transients_central_peak}
  \,.
\end{align}

\begin{figure}[tb]
  \begin{center}
    \includegraphics[width=0.475\textwidth]{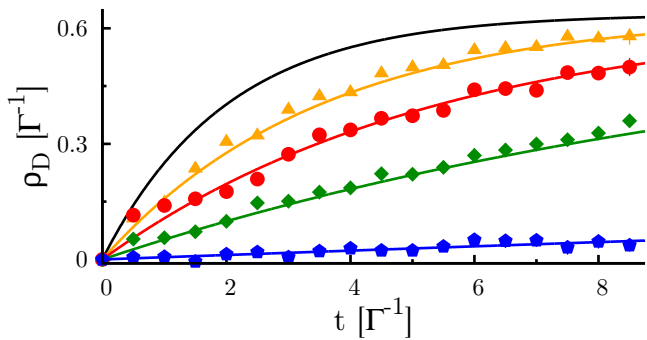}
    \caption{Time-dependent diagMC results for the central resonance at $\omega=0$ with $V=2\Gamma$ for various phonon parameters: $\lambda=3\Gamma$, $\omega_0=4\Gamma$ (yellow triangles), $\lambda=2\Gamma$, $\omega_0=2\Gamma$ (red circles), $\lambda=4\Gamma$, $\omega_0=3\Gamma$ (green diamonds), and $\lambda=8\Gamma$, $\omega_0=4\Gamma$ (blue pentagons). The bandwidth is set to $2 \epsilon_{\text{c}}=8\Gamma$. The corresponding transients in the spirit of Ref.~[\onlinecite{apta}], given by Eq.~(\ref{eq:long_transients_central_peak}), are lines with the same color code as the respective numerical data. The analytical result of the transient in the absence of phonons is shown as a black line. \label{Fig:time_dependent_spectral_function_long_transient}}
  \end{center}
\end{figure}

Our diagMC results confirm this behavior as shown in Fig.~\ref{Fig:time_dependent_spectral_function_long_transient} for various electron-phonon couplings in the moderate polaronic regime. We would like to emphasize the broad range of phonon parameters, for which Eq.~(\ref{eq:long_transients_central_peak}) provides accurate results. Besides confirming the existence of phonon-induced long timescales\cite{apta} determined by $\tilde{\Gamma}^{-1}$, an important consequence of this behavior is that in the low-bias regime the Friedel-Langreth sum rule is fulfilled. That is, the central peak is pinned to $\rho_{\text{D}}(\omega=0)=2/(\pi \Gamma)$ independent of the electron-phonon coupling strength.

In the subsequent discussion, we will study a parameter regime where nonequilibrium effects are most pronounced. According to our preceding discussion, it must therefore be neither close to the limiting cases of a very weak phonon coupling nor in the strong polaronic regime. Another requirement is that we are able to extract the complete LDOS from the time-dependent diagMC results. Therefore, the steady state has to be reached within times which are accessible by diagMC. A reasonable choice of parameters fulfilling these requirements is $\lambda=\omega_0=2\Gamma$: the moderate polaronic regime is accessed and the longest timescales of the transients are roughly $\tilde{\Gamma}^{-1} \approx 2.7 \Gamma^{-1}$. The current implementation of the diagMC is able to simulate up to $t \approx \mathcal{O} (10 \Gamma^{-1})$ within reasonable computational effort. Consequently, we can not only discuss the transient dynamics, but it is also reasonable to extract the steady state of the system from the time-dependent results.

\begin{figure}[tb]
  \begin{center}
    \includegraphics[width=0.475\textwidth]{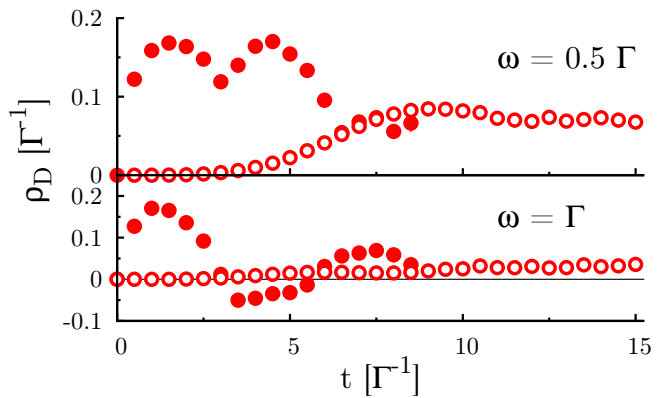}
    \caption{diagMC results for $\rho_{\text{D}}(\omega,t)$ for $\lambda=\omega_0=V=2\Gamma$. The upper panel shows $\omega=0.5\Gamma$, the lower one $\omega=\Gamma$. Empty circles denote a smooth switching within $\tau_{\text{sw}}=12\Gamma^{-1}$, and filled ones correspond to an instantaneous coupling at $t=0$. The bandwidth is $2 \epsilon_{\text{c}}=8\Gamma$.\label{Fig:time_dependent_spectral_function_l2o2_dip}} 
  \end{center}
\end{figure}

In Fig.~\ref{Fig:time_dependent_spectral_function_l2o2_dip} the transients which establish the spectral density at $\omega=0.5\Gamma$ and $\omega=\Gamma$ are shown. Similar to the weak-coupling regime, an instantaneous coupling of the electrodes to the quantum dot leads to an overshooting for small times. The steady state, however, is approached non-monotonically in an oscillating manner. The characteristic timescale for the convergence towards the steady state can be estimated to be given by $\tilde{\Gamma}^{-1}$, pronounced features in the oscillations occur with a periodicity given by $t \approx 2 \pi / \omega_0$. These transients can be reduced by a smooth switch-on procedure of the leads to the quantum dot. If a rather long switch-on time is used, the steady state can be extracted with good accuracy.

\begin{figure}[tb]
  \begin{center}
    \includegraphics[width=0.475\textwidth]{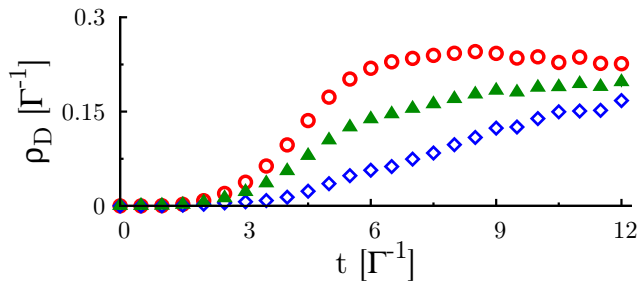}
    \caption{Same plot as in Fig.~\ref{Fig:time_dependent_spectral_function_l2o2_dip}, however, for the first phonon sideband with $\omega= \omega_0= 2 \Gamma$ (red circles) and $\omega= -\omega_0= -2 \Gamma$ (blue diamonds). The green triangles denote the diagMC results of the average $\rho_{\text{D,av}}(\omega,t)$, as in Eq.~(\ref{eq:average_transient_spectral_density}). A smooth switching within $\tau_{\text{sw}}=6\Gamma^{-1}$ is employed. \label{Fig:time_dependent_speck_mum2}}
  \end{center}
\end{figure}

It is worth noticing an interesting effect that is clearly observed in the transient regime for the strong-coupling case: while the particle-hole symmetry requires that the stationary spectral density is symmetric, $\rho_{\text{D}}(\omega)=\rho_{\text{D}}(-\omega)$, this relation doesn't have to be fulfilled necessarily in the transients. In Fig.~\ref{Fig:time_dependent_speck_mum2} an overshooting for a positive frequency, $\omega=\omega_0$, can be observed, whereas the transients for a negative frequency, $\omega=-\omega_0$, show a monotonic increase in time. This effect can be explained by the asymmetry in the initial preparation: right before the coupling of the quantum dot to the leads, the quantum dot is empty. Therefore, only resonances positive in energy exist, since no deexcitation of phonons is possible at $t=0$. On the finite timescale necessary to establish the spectral density \cite{Lothar_spectral_density} this asymmetry in the initial preparation leads to an asymmetry in the transients. Combined with the phonon-induced long timescales, this fact provides a deeper understanding of the splitting of the current depending on the initial preparation on a timescale given by $\mathcal{O}\left(\Gamma^{-1}\right)$, which was observed, e.g., in Ref.~[\onlinecite{bistable}]. This is further corroborated by the fact that for the considered particle-hole symmetric case one obtains $\rho_{\text{D}}^{\text{empty}}(\omega,t) = \rho_{\text{D}}^{\text{occupied}}(-\omega,t)$, where the superscript denotes the initial occupation of the quantum dot.

Since neither $\rho_{\text{D}}(\omega=\omega_0,t)$, nor $\rho_{\text{D}}(\omega=-\omega_0,t)$ exhibit a clear steady state in Fig.~\ref{Fig:time_dependent_speck_mum2}, we use the particle-hole symmetry of the considered setup, and define the average of the transients by

\begin{align}
  \rho_{\text{D,av}}(\omega,t)
  &=
  \frac{1}{2}
  \left[
    \rho_{\text{D}}(\omega,t)
    +
    \rho_{\text{D}}(-\omega,t)
  \right]
  \label{eq:average_transient_spectral_density}
  \,.
\end{align}
The particle-hole symmetry ensures that this function has the same steady-state value as $\rho_{\text{D}}(\omega,t)$ and $\rho_{\text{D}}(-\omega,t)$, separately. The transient dynamics, however, show a quicker convergence towards the plateau value due to the averaging between excitations and deexciations in Eq.~(\ref{eq:average_transient_spectral_density}). Therefore, the steady state of $\rho_{\text{D,av}}(\omega,t)$ can be extracted with reasonable accuracy. We would like to note that similar convergences of the observables can also be found for the currents: While the average current reaches a plateau for $t \gtrsim 8 \Gamma^{-1}$, the left and the right current converge to a joint steady state for times $t \gtrsim 11 \Gamma^{-1}$.

\begin{figure}[tb]
  \begin{center}
    \includegraphics[width=0.475\textwidth]{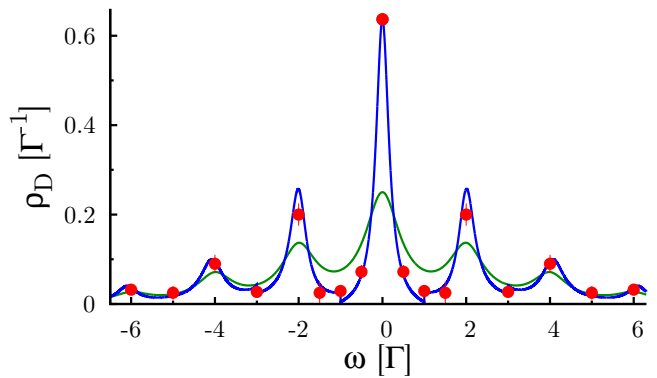}
    \caption{Zero temperature spectral density for $\lambda=\omega_0=2\Gamma$ and $V=2\Gamma$. The results extracted from diagMC are shown with red filled circles, the ISA is a blue line, and SPA is green. \label{Fig:spectral_function_l2o2}}
  \end{center}
\end{figure}

In Fig.~\ref{Fig:spectral_function_l2o2}, we plot the extracted spectral density of the quantum dot, and we make a comparison between diagMC, ISA, as well as SPA. It should be stressed that the error bar of the extracted steady state from the time-dependent diagMC results is twice the total change of $\rho_{\text{D,av}}(\omega, t)$ from $t=8\Gamma^{-1}$ to $t=12\Gamma^{-1}$.

A very sharp central peak at $\omega=0$ is observed with a height given by the Friedel-Langreth sum rule.  Compared to the case of absent phonons, the width of the central peak is reduced to $\tilde{\Gamma} \approx \Gamma e^{-g}$, as it was also observed in Ref.~[\onlinecite{0022-3719-13-24-011}] for the equilibrium situation. This observation eventually proves the close connection between the long transients and phonon narrowing of the resonances, which was discussed in Ref.~[\onlinecite{apta}] by means of an approximative method.

Phonon sidebands can be found at multiples of the phonon frequency, with an exponentially decreasing height. This behavior reflects the fact that transport outside the voltage window is strongly suppressed. Moreover, clear dips in the spectral density appear between two phonon sidebands.

The ISA describes the results obtained from diagMC with remarkably good accuracy. Small differences are only visible at the first phonon sidebands, where the ISA predicts a slightly larger value. Since one basic assump- tion of the ISA is an equilibrium phonon distribution, we conclude that effects due to a (possible) nonequilib- rium phonon distribution only play a minor role in the moderate polaronic regime at low biases.

Regarding the low-frequency domain calculations using the SPA a clear deviation from the diagMC is observed. This means that the charge on the quantum dot is strongly correlated with the excitation of phonons for frequencies that are not too large. In the large frequency domain, however, a good agreement between the diagMC, and the SPA is observed. Therefore, the electrons are decoupled from the phonons for transport with energies much larger than the voltage window, which confirms the results of Ref.~[\onlinecite{carmina2009}].

\subsection{Far-from-equilibrium spectral density}

\begin{figure}[t]
  \begin{center}
    \includegraphics[width=0.475\textwidth]{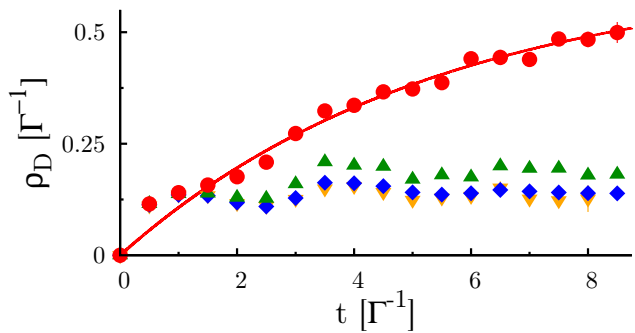}
    \caption{Time-dependent diagMC results for the transients, which establish the central peak $\rho_{\text{D}}(\omega=0)$ for $\lambda=\omega_0=2\Gamma$ and various voltages: $V=2\Gamma$ (red circles, bandwidth $2 \epsilon_{\text{c}}=8\Gamma$), $V=6\Gamma$ (green triangles, $2 \epsilon_{\text{c}}=8\Gamma$), $V=10\Gamma$ (blue diamonds, $2 \epsilon_{\text{c}}=12\Gamma$), $V=14\Gamma$ (yellow reverted triangles, $2 \epsilon_{\text{c}}=16\Gamma$). The approximative description in the spirit of Ref.~[\onlinecite{apta}], given by Eq.~(\ref{eq:long_transients_central_peak}), is shown as a red line. \label{Fig:time_dependent_spectral_function_l2o2_voltage}}
  \end{center}
\end{figure}

In the subsequent discussion, we will analyze nonequilibrium effects in the moderate polaronic regime. For this purpose we consider voltage windows that contain one or more phonon sidebands. While nonequilibrium effects are not important in the weak-coupling regime, as discussed in Section~\ref{electronic_regime}, the effect of a large bias voltage is strong in the moderate polaronic regime: The transients, which establish the central transport channel at $\omega=0$, are shown in Fig.~\ref{Fig:time_dependent_spectral_function_l2o2_voltage}, where the same phonon parameters are used as in the previous section: $\lambda=\omega_0=2\Gamma$. For a voltage window, which includes one or more phonon sidebands, the transients no longer follow the exponential convergence, given by Eq.~(\ref{eq:long_transients_central_peak}). Moreover, the relevant timescales for the transient dynamics are significantly smaller so that the steady state is reached faster than in the low-bias regime. Furthermore, the steady-state value drops to a much smaller value, which reflects the fact that the transport through the quantum dot outside the low-voltage regime is dominated by the excitation of one or more phonons. This behavior clearly violates the Friedel-Langreth sum rule.

The strongest voltage dependence of the height of the spectral density at $\omega=0$ is observed when increasing the voltage from $V=2\Gamma$ to $V=6\Gamma$. The reason for this behavior is that for $\lambda=\omega_0=2\Gamma$ the central transport channel as well as the first phonon sideband are most pronounced for $V=2\Gamma$ as it can be seen in Fig.~\ref{Fig:spectral_function_l2o2}. Including the first sideband into the voltage window by setting $V=6\Gamma$, charge transport involving single-phonon processes becomes likely and thus this important transport channel opens. This reduces the probability for charge transport without exciting a phonon and, consequently, the height of the central resonance at $\omega=0$ decreases. A further increase of the voltage does not exhibit this pronounced behavior since the transport channels involving two or more phonons are much smaller for the considered parameters.

\begin{figure}[tb]
  \begin{center}
    \includegraphics[width=0.475\textwidth]{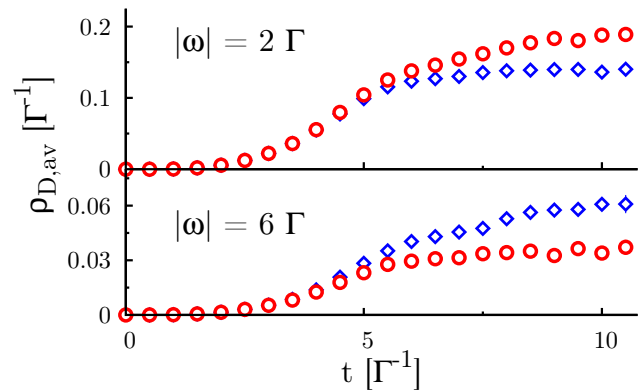}
    \caption{Time-dependent diagMC results for $\rho_{\text{D},\text{av}}(\omega,t)$ with $|\omega| = \omega_0$ (upper panel) and $|\omega| = 3 \omega_0$ (lower panel) with $\lambda=\omega_0=2\Gamma$. $V=2\Gamma$ is highlighted with red circles (bandwidth $2 \epsilon_{\text{c}}=8 \Gamma$ in the upper panel, and $2 \epsilon_{\text{c}}=14 \Gamma$ in the lower one), and $V=10\Gamma$ with blue diamonds (bandwidth $2 \epsilon_{\text{c}}=14 \Gamma$ for both panels). A smooth switching within $\tau_{\text{sw}}=6 \Gamma^{-1}$ is employed. \label{Fig:time_dependent_spectral_function_l2o2_v10_sideband}}
  \end{center}
\end{figure}

An important consequence of the decreasing central transport channel is that the weight of the LDOS is shifted towards larger frequencies. A similar shift has been reported recently in the differential conductance \cite{PhysRevB.87.195112}. This shift causes an increase of the phonon sidebands outside the voltage window as it can be seen in the lower panel of Fig.~\ref{Fig:time_dependent_spectral_function_l2o2_v10_sideband}. Including a phonon sideband into the voltage window causes a drop of the phonon resonance as it can be seen in the upper panel of Fig.~\ref{Fig:time_dependent_spectral_function_l2o2_v10_sideband}.

\begin{figure}[tb]
  \begin{center}
    \includegraphics[width=0.46\textwidth]{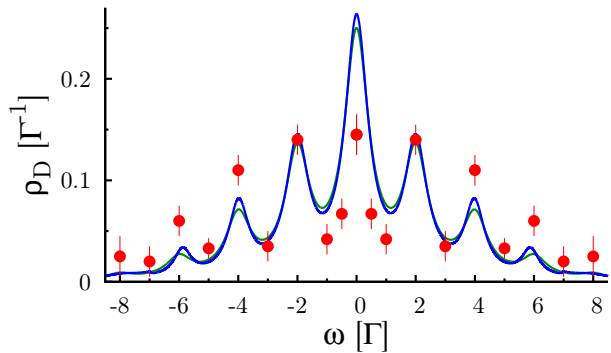}
    \caption{Same color code as Fig.~\ref{Fig:spectral_function_l2o2} but with $V=10\Gamma$.\label{Fig:spectral_function_l2o2_v10}}
  \end{center}
\end{figure}

The resulting spectral density extracted from the time-dependent diagMC for $V=10\Gamma$ is shown in Fig.~\ref{Fig:spectral_function_l2o2_v10}. A comparison with the low-bias LDOS in Fig.~\ref{Fig:spectral_function_l2o2} reveals that inside the voltage window all peaks seem to align to a similar height, whereas peaks outside the voltage window are increased. Moreover, the narrowing of the central transport channel to $\tilde{\Gamma} \approx \e^{-g} \Gamma$ is no longer observed in the large voltage regime. Rather, a width of approximately $\Gamma$, which is the value for the interaction-free case, is recovered. We note, however, that this is not an indication that the charge transport through the quantum dot is uncorrelated from the phonons: a comparison with the SPA reveals that the central transport channel is strongly suppressed.

In addition, Fig.~\ref{Fig:spectral_function_l2o2_v10} reveals a clear deviation of the ISA from the diagMC results. For such a large voltage the ISA spectral density has (almost) converged to the SPA case. It is interesting to remark that a recent diagrammatic resummation scheme valid in the polaronic regime \cite{ruben} predicts a similar convergence towards the SPA for large voltages. As these approximate theories do not include the effect of the nonequilibrium distribution of phonons in a self-consistent way, the numerically exact diagMC results strongly indicate that this effect is important in the moderate polaronic regime in the large bias limit.

A straightforward way to confirm that the deviations observed in Fig.~\ref{Fig:spectral_function_l2o2_v10} are indeed produced by a nonequilibrium phonon distribution is to try to simulate the results by an equilibrium distribution at some effective temperature. For this purpose, we use the SPA since the ISA as well as the approach of Ref.~\onlinecite{ruben} converge to the SPA for sufficiently large bias voltages.

\begin{figure}[t]
  \begin{center}
    \includegraphics[width=0.46\textwidth]{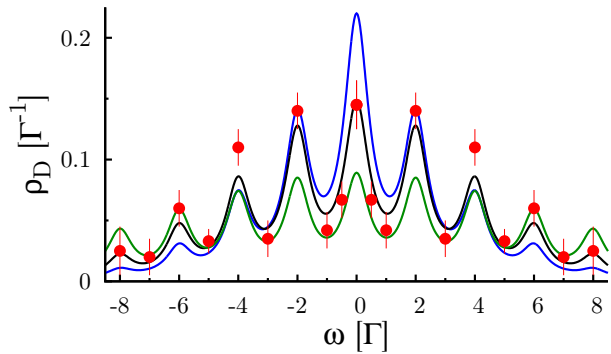}
    \caption{Same diagMC results as in Fig.~\ref{Fig:spectral_function_l2o2_v10}. The SPA has been calculated with an effective temperatures of the phonons: $T= \Gamma$ (blue), $T=3 \Gamma$ (black), and $T=10 \Gamma$ (green).\label{Fig:spectral_function_l2o2_v10_temp}}
  \end{center}
\end{figure}

In Fig.~\ref{Fig:spectral_function_l2o2_v10_temp} the SPA results given by Eq.~(\ref{eq:spectral_function_spa}) for various effective phonon temperatures $T \to T_{\text{eff}}$ are compared to the diagMC results, which are calculated for $V=10 \Gamma$ and $T=0$. Strikingly, the central peak of the SPA spectral density decreases by increasing the phonon temperature, whereas phonon resonances well away from the central peak increase -- a behavior similar to the diagMC results for increasing voltage. For an effective phonon temperature of roughly $T_{\text{eff}} \approx 3 \Gamma$, the results from SPA match the diagMC results for $T=0$ and $V=10\Gamma$ with good accuracy. We would like to note that despite the good overall agreement, small differences can be observed for the second phonon resonance, which means that not all effects can be completely described in detail by this effective theory. 


\section{Conclusions}

In this paper we calculated the nonequilibrium spectral density of a vibrational quantum dot using the numerical exact diagMC technique and compared its predictions with those of approximate methods such as ISA and SPA.

We showed that for a weak electron-phonon interaction the spectral density of the quantum dot resembles the equilibrium one independently of the bias voltage. For intermediate electron-phonon coupling strengths in the moderate polaronic regime, we determined a significant voltage dependence of the spectral density. An increasing bias voltage shifts the weight of the spectral density towards larger energies: The central resonance decreases, whereas phonon resonances outside the voltage window are increased with respect to the equilibrium results. Inside the voltage window our results indicate that the phonon peaks align to a similar height. We were able to link the voltage dependence of the spectral density to an effective ``heating'' of the phonons caused by inelastic excitations.

The explicit voltage dependence of the spectral density points out the importance of accessing the spectral density directly, e.g., by means of a three-terminal setup \cite{sun_guo_third_terminal2001,lebanon_schiller_third_terminal2001,leturcq_three_terminal2005}. An indirect measurement, e.g., of the differential conductance might lead to a discrepancy between the result and the actual spectral density due to its voltage dependence.

Another consequence of our findings is that for future descriptions by means of approximative approaches it is desirable to also account for nonequilibrium effects of the phonon distribution, which could be preformed, e.g., as proposed in Ref.~[\onlinecite{Galperin06}].

Finally, we would like to emphasize that for small voltages, we confirmed the existence of phonon-induced long transients previously proposed in Ref.~[\onlinecite{apta}]. Moreover, we pointed out that the inverse of the width of the resonances in the spectral density determines the relevant timescale in the system.

\vspace{-0.5cm}

\section*{Acknowledgments}

The authors like to thank R. C. Monreal, A. Levy Yeyati, R. Seoane Souto, and A. Komnik for many fruitful discussions. KFA acknowledges the computing time at the bwGRID and Juropa in J\"{u}lich. This work was financially supported by Spanish Mineco through grant FIS2011-26516, and by the NSF (PIF-1211914 and PFC-1125844).


%
\bibliographystyle{apsrevM} \bibliography{papers}

\end{document}